%% file: amai.tex
\pdfoutput=1
\RequirePackage{amsmath}
\documentclass[smallextended]{svjour3}
\usepackage[T1]{fontenc}
\usepackage{microtype}
\usepackage{url}
\usepackage[hidelinks]{hyperref}
\usepackage{mathtools}
\usepackage{amssymb}
\usepackage[bf,small]{caption}
\usepackage{tikz}
\usetikzlibrary{positioning}

\title{The SAT+CAS Method for Combinatorial Search with Applications to Best Matrices}

\author{Curtis Bright\and 
Dragomir \v{Z}. \DJ{}okovi\'c\and 
Ilias Kotsireas\and 
Vijay Ganesh 
}

\journalname{Annals of Mathematics and Artificial Intelligence}
\date{}

\institute{C.~Bright, D.~\DJ{}okovi\'c, V.~Ganesh \at University of Waterloo \\
200 University Ave W \\
Waterloo, Ontario, Canada
\and
I.~Kotsireas \at Wilfrid Laurier University \\
75 University Ave W \\
Waterloo, Ontario, Canada}

\authorrunning{C.~Bright, D.~\DJ{}okovi\'c, I.~Kotsireas, V.~Ganesh}
\titlerunning{The SAT+CAS Method for Combinatorial Search}

\newcommand\elide{[\,\dots]}
\newcommand\A{\bar A}
\newcommand\B{\bar B}
\newcommand\C{\bar C}
\newcommand\D{\bar D}
\newcommand\Z{\mathbb Z}
\newcommand\MC{\textsc{MathCheck}}
\DeclareMathOperator{\s}{sum}
\DeclareMathOperator{\PSD}{PSD}
\DeclarePairedDelimiter{\abs}{\lvert}{\rvert}
\DeclarePairedDelimiter{\paren}{\lparen}{\rparen}
\DeclarePairedDelimiter{\floor}{\lfloor}{\rfloor}
\newcommand\p{{\tt+}}
\newcommand\m{{\tt-}}
\smartqed

\begin{document}

\maketitle

\setcounter{footnote}{0}

\begin{abstract}

  In this paper, we provide an overview of the SAT+CAS method that
  combines satisfiability checkers (SAT solvers) and computer algebra
  systems (CAS) to resolve combinatorial conjectures, and present new
  results vis-\`a-vis best matrices.  The SAT+CAS method is a variant
  of the Davis--Putnam--Logemann--Loveland
  DPLL($T$) architecture, where the $T$ solver is replaced by a CAS.
  We describe how the SAT+CAS method has been previously used to resolve many
  open problems from graph theory, combinatorial design theory, and
  number theory, showing that the method has broad applications across
  a variety of fields.  Additionally, we apply the method to construct the
  largest best matrices yet known and present new skew Hadamard matrices
  constructed from best matrices.  We show the best matrix conjecture (that
  best matrices exist in all orders of the form $r^2+r+1$) which was
  previously known to hold for $r\leq6$ also holds for $r=7$.
  We also confirmed the results of the exhaustive searches that have
  been previously completed for $r\leq6$.
  
\end{abstract}

\section{Introduction}\label{sec:introduction}

In recent years a number of search paradigms have emerged that allow
the solving of extraordinarily large problems in combinatorial
mathematics. In particular, one of the most successful techniques has been
the ``SAT paradigm'' of reducing a problem into Boolean logic and then
searching for a solution using a SAT solver~\cite{handbook}.
In fact, the SAT paradigm has been so successful that it is routinely
used to solve problems in areas of mathematics which don't seem to be
directly connected to Boolean logic at first glance.  In 2017, Heule,
Kullmann, and Marek~\cite{heule2017solving} summarized the
state-of-the-art in combinatorial searches as follows:

\begin{quote}
Surprisingly, SAT solving is getting so strong that indeed [using SAT
  solvers] seems today the best solution in most cases.
\end{quote}

Some enormous combinatorial problems have been resolved in this way.
In particular, the cube-and-conquer SAT solving
paradigm~\cite{heule2017science} by Heule and Kullmann has achieved a
number of striking successes including solving the Boolean Pythagorean
triples problem~\cite{heule2016solving} and determining the value of
the fifth Schur number---a problem that resisted solution for over 100
years~\cite{heule2018schur}.

Briefly, in the cube-and-conquer paradigm a ``look-ahead'' SAT
solver~\cite{heule2009look} partitions the search space into a number
of independent subspaces of roughly equal difficulty called ``cubes''.
Each cube is then solved by a SAT solver using conflict-driven
clause learning~\cite{marques2009conflict} (possibly employing
parallelization across a large number of processors) to determine if a
solution to the problem exists.  See Section~\ref{sec:background} for
more background on the cube-and-conquer paradigm.

Despite these impressive successes, the cube-and-conquer paradigm is
not appropriate for all kinds of combinatorial problems, and in
particular it would be difficult to use in problems that have
properties that \emph{cannot easily be expressed in Boolean logic}.
When dealing with such problems one common approach is to employ an
SMT (SAT modulo theories)
solver~\cite{handbook-smt,z3,yices,ganesh-stp} based on the
Davis--Putnam--Logemann--Loveland algorithm (modulo theories) denoted
by DPLL($T$)~\cite{not06} where $T$ is a theory of first-order logic.
SMT solvers can solve many problems of interest in the context of
automatic program verification~\cite{botinvcan2009separation} and
automatic test case generation~\cite{cadar2008exe}.  However, modern
SMT solvers only support specific theories and in this paper we are
interested in solving combinatorial problems that have properties that
cannot be easily expressed in those theories.

By contrast, a huge number of mathematical properties of interest can
easily be expressed in \emph{computer algebra systems} (CAS) such as
Maple~\cite{mapleguide}, Mathematica~\cite{wolfram2003mathematica},
and SageMath~\cite{sagemath}.  Indeed, a common approach for solving
problems that use advanced mathematics is to write a program in the
programming language of a CAS.  While CAS are very impressive in
solving pure math problems, they are not optimized for combinatorial
problems that require both math and search.

We therefore have developed a new ``SAT+CAS'' method, that combines the
best of both the SAT world (for search) and CAS (for math) and have used this
method to solve several large combinatorial problems that rely on advanced
mathematics. We do this by using a combination of SAT solvers and
computer algebra systems and use each system in ways that exploit its strengths.
Namely, we use the SAT solver as the combinatorial search engine and use the
CAS to check properties that are too difficult or cumbersome to encode into
Boolean logic. The SAT+CAS paradigm is a variant of DPLL($T$) and can be
captured as DPLL(CAS), however, it also uses the CAS more generally.
For example, during preprocessing the CAS can often
show the equivalence of subproblems that the SAT solver would
not be able to discern and would therefore
otherwise have to solve multiple times.

As concrete examples of this paradigm we mention three problems and
the properties that we checked using a CAS\@.  See
Section~\ref{sec:preliminaries} for background on these problems and
Section~\ref{sec:sat+cas} for details of how we used the SAT+CAS
method to push the state-of-the-art in these problems. We briefly
mention these below:

\begin{enumerate}
\item The Ruskey--Savage conjecture
  (see~\cite{zulkoski2017combining,zulkoski2015mathcheck}).  This
  conjecture states that any matching of a hypercube graph can be
  extended to a Hamiltonian cycle. We encode the property that a set
  of edges is a matching in a SAT instance and check that the edges
  extend to a Hamiltonian cycle using a CAS.

\item Enumerating Williamson matrices
  (see~\cite{bright2018sat,bright2018satcas}).  Williamson matrices
  are square $\{\pm1\}$-matrices that satisfy a simple arithmetical
  property. They also satisfy a more complicated property based on the
  discrete Fourier transform that we check using a CAS.

\item Enumerating complex Golay pairs
  (see~\cite{bright2018enumeration,bright2018complex}).  Complex Golay
  pairs are two polynomials with coefficients in $\{\pm1,\pm i\}$ that
  satisfy a simple arithmetical property. The norm of the polynomials
  satisfy certain bounds that we check using the nonlinear programming
  optimizer of a CAS.
\end{enumerate}

\subsection{New results}

In this paper, we further extend the success of the SAT+CAS paradigm
to another class of matrices studied in combinatorial design theory
known as best matrices~\cite{georgiou2001circulant}.  This case study
is similar to the Williamson example from above because best matrices
are also known to satisfy a strict condition based on the discrete
Fourier transform.  However, best matrices tend to be much rarer than
Williamson matrices.  In particular, it is known that if circulant
best matrices of order $n$ exist then $n$ must be of the form
$r^2+r+1$ for some $r\geq0$~\cite{djokovic2018goethals}.

Before this work it was known that best matrices exist in all these
orders for~$r$ up to and including $r=5$~\cite{georgiou2001circulant}
and best matrices were recently found for
$r=6$~\cite{djokovic2018goethals}.  This makes it tempting to
conjecture that best matrices actually exist for all orders of the
form $r^2+r+1$.  See Section~\ref{sec:best} for a detailed discussion
of how we applied the SAT+CAS paradigm to this problem and
Section~\ref{sec:implementation} for details on our implementation and
results.

The main new result of this paper is
that we show for the first time that best matrices exist for $r=7$
by explicitly constructing best matrices of order $57$---the largest
best matrices currently known. Additionally, we use these
matrices to construct new skew Hadamard matrices and perform the first
published verification of the counts of best matrices given
in~\cite{djokovic2018goethals,georgiou2001circulant} (see
Section~\ref{sec:implementation}). A secondary contribution is a
demonstration that the SAT+CAS method is applicable to a wide variety
of fields including graph theory, combinatorial design theory, and
number theory (see Section~\ref{sec:sat+cas}).

\section{Previous work}\label{sec:background}

SAT solving has been studied since the 1960s but it was not generally
considered a tractable problem at that time.  In fact, SAT was often
considered as a typical example of an \emph{in}tractable problem after
the Cook--Levin theorem showed that SAT is NP-complete.
In the 1990s new algorithmic techniques such as
conflict-driven clause learning and
variable branching heuristics greatly increased the size of problems
that could be solved.  This spurred what has been referred to as a
``SAT revolution''~\cite{vardi2009symbolic} and SAT solvers now
routinely solve problems with millions of variables and constraints.
The ability to efficiently solve such problems has resulted in
a large number of applications
of SAT solving ranging from formal verification of
hardware~\cite{vizel2015boolean} to solving Sudoku
puzzles~\cite{lynce2006sudoku}.  In this paper we focus on
applications of SAT solvers to combinatorial search problems.

The first application of SAT solvers to combinatorial problems appears
to be by McCune~\cite{mccune1994davis}, and Stickel and
Zhang~\cite{stickel1994first} who in the 1990s used SAT solvers to
solve a number of open Latin square and quasigroup problems.  Zhang
developed a solver called SATO~\cite{zhang1997specifying} and applied
it to numerous Latin square problems and observed that such an
approach was just as effective as using a special-purpose
solver~\cite{zhang2009combinatorial}:

\begin{quote}
In the earlier stage of our study of Latin square problems, the author
wrote two special-purpose programs.  After observing that these two
programs could not do better than SATO, the author has not written any
special-purpose search programs since then.
\end{quote}

In the 2000s SAT solvers were also successfully applied to the branch
of combinatorics known as Ramsey theory and in particular to the
problem of computing van der Waerden numbers.  The mathematician van
der Waerden proved~\cite{van1927beweis} that any $r$-colouring of the
natural numbers must contain~$k$ numbers in arithmetic progression
that are all the same colour (monochromatic).  A \emph{van der Waerden
  number} is the smallest value of~$n$ such that all $r$-colourings of
$\{1,\dotsc,n\}$ have a monochromatic arithmetic progression of length
$k$.

An initial result in 2003 was by Dransfield, Marek, and
Truszczy{\'n}ski~\cite{dransfield2003satisfiability} who used a SAT
solver to significantly improve the lower bounds on several van der
Waerden numbers.  Two years later Kouril and
Franco~\cite{kouril2005resolution} used a SAT solver to find a
$2$-colouring of $\{1,\dotsc,1131\}$ without any monochromatic
arithmetic progressions of length~$6$ and conjectured that it was not
possible to increase the size of this set.  Three years later Kouril
and Paul~\cite{kouril2008van} used a SAT solver to prove this, in
other words they showed that all $2$-colourings of $\{1,\dotsc,1132\}$
contain monochromatic arithmetic progressions of length~$6$.

In 2011, Heule, Kullmann, Wieringa, and Biere~\cite{heule2011cube}
developed the cube-and-conquer paradigm in the process of solving SAT
instances that arose from computing van der Waerden
numbers~\cite{ahmed2014van}.  They found that the cube-and-conquer
method performed better than any other method on these instances:

\begin{quote}
Results on hard van der Waerden benchmarks using our basic method show
reduced computational costs up to a factor 20 compared to the fastest
``pure'' SAT solver.
\end{quote}

The basic idea behind the cube-and-conquer method is to combine two
different SAT solving strategies, the ``lookahead'' and
``conflict-driven'' strategies.  Lookahead solvers are good at making
decisions at a \emph{global} level, i.e., finding the next decision
that simplifies the problem as much as possible.  In contrast,
conflict-driven solvers are good at solving large problems that admit
a relatively short solution, i.e., ones that can be solved by making
specific \emph{local} decisions that may not be good globally but
happen to work in that specific case.

The crucial insight by Heule et al.\ is to employ lookahead solvers to
split the problem into many subproblems and then switch to
conflict-driven solvers once the subproblems become simple enough.  In
this way a cube-and-conquer solver performs better than either a pure
lookahead or pure conflict-driven solver.  Furthermore the method
naturally admits parallelization as the subproblems can be solved
using separate processors.

The cube-and-conquer paradigm has been enormously successful.  In
particular, Heule, Kullmann, and Marek~\cite{heule2016solving} used it
to solve the Boolean Pythagorean triples problem and
Heule~\cite{heule2018schur}
used it to find the value of the fifth Schur number---both
of these problems were well-known and went unsolved for decades.
SAT solvers have also been used to compute Green--Tao numbers by
Kullmann~\cite{kullmann2010green} and solve a special case of the
Erd\H os discrepancy conjecture by Konev and
Lisitsa~\cite{konev2014sat}.

\section{Mathematical preliminaries}\label{sec:preliminaries}

In this section we describe the mathematical preliminaries necessary to
understand the problems discussed in this paper.  One of our goals
is to demonstrate that the SAT+CAS method is applicable across many fields,
so to this end we introduce problems in
the fields of graph theory, combinatorial design theory, and number theory.

\paragraph{Graph theory.}
The \emph{hypercube graph} of order~$n$ is a graph on $2^n$ vertices where
the vertices are labelled with bitstrings of length~$n$.  Two vertices
are adjacent in this graph exactly when their labels differ in a single bit.
A \emph{matching} of a graph is a subset of its edges such that no two
edges share a vertex.  A \emph{Hamiltonian cycle} of a graph is a path
through the graph that starts and ends at the same vertex and visits each
vertex exactly once.

The Ruskey--Savage conjecture says that every matching of a hypercube
graph of order $n\geq2$ can be extended into a Hamiltonian cycle of the graph~\cite{ruskey1993hamilton}.
This conjecture has been open for over twenty-five years.

\paragraph{Combinatorial design theory.}
A \emph{Hadamard matrix} is a square matrix with $\pm1$ entries such that
any two distinct rows are orthogonal.  Hadamard matrices have a long
history (first constructed in 1867~\cite{sylvester1867thoughts}) and applications to error-correcting
codes~\cite{macwilliams1977theory}, image coding~\cite{pratt1969hadamard}, and techniques for statistical estimation~\cite{rao1996balanced}.
The \emph{Hadamard conjecture}
is that Hadamard matrices exist in all orders that are multiples of four
and much work has gone into constructing Hadamard matrices in as many orders
as possible~\cite{hedayat1978hadamard,horadam2012hadamard,kharaghani2005hadamard,seberry1992hadamard}.
One way of constructing a Hadamard matrix of order~$4n$
is to use a set of Williamson matrices of order~$n$~\cite{kotsireas2009hadamard}.

To define Williamson matrices we require the definition of a circulant matrix.
A matrix is \emph{circulant} if each row is equal to the previous row
shifted by one element to the right (with a wrap-around).  Therefore, a circulant
matrix can equivalently be identified with the sequence formed by its first row.
Four circulant and symmetric matrices $A$, $B$, $C$, $D$ of order $n$ are \emph{Williamson matrices}
if they have $\pm1$ entries and $A^2+B^2+C^2+D^2$ is the scalar matrix $4nI$.
The \emph{Williamson conjecture} is that Williamson matrices exist in all positive orders $n$~\cite{golomb1963search}
and the \emph{even Williamson conjecture} is that Williamson matrices exist in all even orders~\cite{bright2019applying}.

Best matrices are similar to Williamson matrices and will be formally
defined in Section~\ref{sec:best}.  One of the major differences is that
best matrices can be used to construct \emph{skew} Hadamard matrices,
i.e., ones whose off-diagonal entries are anti-symmetric.
Much effort has also been spent constructing skew Hadamard matrices
in as many orders as possible.  The \emph{skew Hadamard conjecture} says
that they exist in all orders of the form $4n$~\cite{craigen2007hadamard} but
the current smallest unknown order is $n=69$ and the previous
smallest unknown order $n=47$ was solved in 2008~\cite{djokovic2007skew}.

Best matrices are primarily used to construct orthogonal designs and skew Hadamard matrices---%
which have applications to fields such as statistical analysis and coding theory.
For example, Kim and Sol\'e~\cite{kim2008skew} have shown that skew Hadamard
matrices of order~$4n$ produce self-dual codes over fields of characteristic dividing~$n$.
Applications to other combinatorial structures relying
on best matrices are given in the original paper of
Georgiou, Koukouvinos, and Seberry that defined best matrices~\cite{georgiou2001circulant}.
For example, they use best matrices to construct the first
known orthogonal designs of a certain form.

\paragraph{Number theory.}
Two polynomials $A$ and $B$ with coefficients in $\{\pm1,\pm i\}$ and of
length $n$ (i.e., degree $n-1$)
form a \emph{complex Golay pair} if $\abs{A(z)}^2+\abs{B(z)}^2=2n$
for all $z$ on the unit circle.
Such polynomials (with real coefficients) were first used by Golay
to solve a problem in infrared multislit spectrometry~\cite{golay1949multi}.
They have since been applied
to an enormous number of applications in engineering (particularly
in communications~\cite{taghavi2007autocorrelations}).
They also provide extremal examples
for various problems in number theory~\cite{Borwein2002}.

Craigen, Holzmann, and Kharaghani~\cite{craigen2002complex} study complex Golay pairs
and make a number of conjectures concerning them.  For example, they conjecture
that if complex Golay pairs of prime length~$p$ do not exist then complex Golay pairs
also do not exist in any length that is divisible by~$p$.
Thus, since complex Golay pairs of length~$7$ do not exist,
their conjecture implies that complex Golay pairs also do not exist in length~$28$.
Additionally, they conjecture that complex Golay pairs do not exist in length~$23$
based on a partial search.

\section{The SAT+CAS paradigm}\label{sec:sat+cas}

It is well known that one of the drawbacks of Boolean logic is that it
is not expressive enough for many domains~\cite{cimatti2006building}.
This was a significant
impetus for the development of \emph{SAT modulo theories} (SMT)
solvers and the DPLL($T$) architecture~\cite{ganzinger2004dpll} that
can solve problems specified in more expressive theories.  A few SMT
solvers have the ability to work with more mathematically complex
theories such as non-linear transcendental
arithmetic~\cite{cimatti2018incremental}.
However, to our knowledge no SMT solvers can compute
fast Fourier transforms or are optimized to handle the many fragments of
mathematics supported by computer algebra systems.

Conversely, \emph{computer algebra systems} (CAS) from the field of
symbolic computation are optimized to solve hard non-linear algebraic
problems, among many other fragments of mathematics. In 2015,
\'Abrah\'am~\cite{abraham2015building} pointed out that the fields of
symbolic computation and SMT solving have similar aims but the fields
have developed mostly independently of each other:

\begin{quote}
The research areas of SMT solving and symbolic computation are quite
disconnected. On the one hand, SMT solving \elide\ makes use of
symbolic computation results only in a rather naive way.  \elide\ On
the other hand, symbolic computation \elide\ does not exploit the
achievements in SMT solving for efficiently handling logical
fragments, using heuristics and learning to speed-up the search for
satisfying solutions.
\end{quote}

Furthermore, she made the case that these communities could mutually
benefit from exploiting the achievements of the other field.  To this
end the SC$^2$ project (for symbolic computation and satisfiability
checking) was started to bridge the gap between these
communities~\cite{abraham2016sc2}.

Independently of the work of \'Abrah\'am, we started developing a
system called \MC\ in 2014 inspired by the DPLL($T$) algorithm but
replacing the theory solver with a computer algebra system.
We used \MC\ to show that certain graph theoretic conjectures held
up to bounds that had not previously been verified~\cite{zulkoski2015mathcheck}.
Later we applied \MC\ to find (or disprove the existence of)
Williamson matrices~\cite{bright2018satcas}, complex
Golay pairs~\cite{bright2018enumeration}, and good
matrices~\cite{bright2019good} in certain orders.
A summary of these results is presented in Table~\ref{tbl:summary}.

\begin{table}\centering
\begin{tabular}{c@{\qquad}c@{\qquad}c}
\bf Paper & \bf Conjecture & \bf Main Result \\
Bright et al.~\cite{bright2019good} & Good Matrix & Found three new counterexamples \\
Bright et al.~\cite{bright2016mathcheck} & Williamson & Found the smallest counterexample \\
Bright et al.~\cite{bright2019applying} & Even Williamson & Verified orders up to 70 \\
Bright et al.~\cite{bright2018complex} & Complex Golay & Nonexistence in lengths 23, 25, 27, 28 \\
Zulkoski et al.~\cite{zulkoski2015mathcheck} & Ruskey--Savage & Verified in order five \\
Zulkoski et al.~\cite{zulkoski2015mathcheck} & Norine & Verified in order six \\
This work & Best Matrix & Verified in order 57
\end{tabular}
\caption{A summary of the results produced by the SAT+CAS system \MC.}\label{tbl:summary}
\end{table}

The SAT instances that were generated in the graph theory case studies
were small enough that the instances could be solved without splitting
them into subproblems.  However, in every subsequent problem we found
that splitting was essential in order to solve the largest cases.  We
observed very similar behaviour to what was described by
Heule~\cite{heule2018schur} in the cube-and-conquer paradigm, namely,
that instances could be solved much quicker once they had been split
into independent subproblems.  This held true even if only using a
single processor and even when accounting for the time it took to
split the problem.

\begin{figure}
\begin{center}
\begin{tikzpicture}[align=center,node distance=4em]
\node(input){Conjecture and an order~$n$ to verify};
\node[below=of input,text width=12em,minimum height=3em,rectangle,draw](divide2){Generate SAT instances};
\node[node distance=7em,right=of divide2,text width=7em,minimum height=3em,rectangle,draw](cas2){CAS};
\node[below=of divide2,text width=12em,minimum height=3em,rectangle,draw](sat){SAT solver};
\node[below=of cas2,text width=7em,minimum height=3em,rectangle,draw](cas3){CAS};
\node[below=of sat,text width=15em](output){\clap{Verification or counterexample}\\of conjecture in order~$n$};
\node[left of=input,node distance=12em](inputlabel){Input:};
\node[left of=divide2,node distance=12em](preproclabel){Divide:};
\node[left of=sat,node distance=12em](conquerlabel){Conquer:};
\node[left of=output,node distance=12em](outputlabel){Output:};
\draw[->](input)--(divide2);
\draw[->](divide2)--(sat);
\draw[->,transform canvas={yshift=0.5em}](divide2)--node[above,text width=6em]{instances}(cas2);
\draw[<-,transform canvas={yshift=-0.5em}](divide2)--node[below,text width=6em]{inequivalent instances}(cas2);
\draw[->,transform canvas={yshift=0.5em}](sat)--node[above,text width=6em]{\clap{partial satisfying}\\assignment}(cas3);
\draw[<-,transform canvas={yshift=-0.5em}](sat)--node[below,text width=6em]{conflict clause}(cas3);
\draw[->](sat)--(output);
\end{tikzpicture}
\end{center}
\caption{An general overview of the SAT+CAS paradigm using divide and conquer.}\label{fig:sat+cas}
\end{figure}
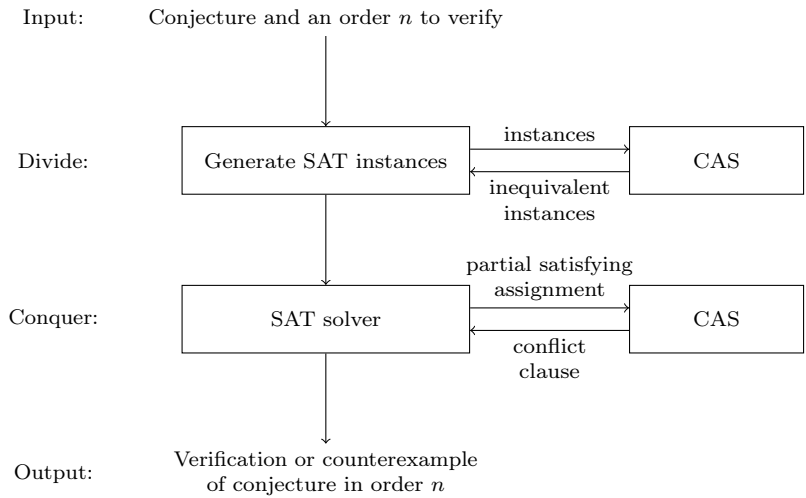

Our SAT+CAS method can be viewed as a special case of the
cube-and-conquer paradigm with two major differences.  First, we don't
divide the problem into subproblems specified by cubes (a conjunction
of literals).  Instead, we specify the division of the problem into subproblems
by using clauses in conjunctive normal form.  Second, we use a computer algebra
system during both the divide and conquer phases.  During the dividing
phase the computer algebra system can often discard entire subproblems
without even sending them to a SAT solver.  For example, if the CAS
finds that two subproblems are isomorphic then one can safely be
discarded.  However, devising a splitting method that takes advantage
of the computer algebra system and performs well requires significant
knowledge of the domain.  We display a general overview of the
SAT+CAS method using divide and conquer in Figure~\ref{fig:sat+cas}.

We now describe in detail the problems outlined in
Sections~\ref{sec:introduction} and~\ref{sec:preliminaries} and how we
applied the SAT+CAS paradigm to derive new results about each
problem. We leverage some of these ideas in Section~\ref{sec:best}
and use them to construct the largest best matrices currently known.

\paragraph{Graph theory.}
The Ruskey--Savage conjecture (that every matching of a hypercube can
be extended into a Hamiltonian cycle) was previously known to hold in the
orders $n=2$, $3$, and~$4$~\cite{fink2007perfect}.  Using \MC\ we showed
that the conjecture also holds in the order $n=5$ for the
first time~\cite{zulkoski2015mathcheck}.

The constraint that a subset of the edges of a hypergraph forms a
matching is encoded directly into Boolean logic.  However, the
constraint that says that such a matching can be extended into a
Hamiltonian cycle is not straightforward to encode in Boolean
logic---but a computer algebra system can easily test this.
Therefore, whenever a satisfying assignment of the propositional
constraints (i.e., a matching of the graph) is found by the SAT
solver the matching is passed to a CAS to verify that it can be extended
into a Hamiltonian cycle.

If the CAS finds that the given matching can not extend to a
Hamiltonian cycle this provides a counterexample of the conjecture.
Otherwise, the CAS provides to the SAT solver a clause that blocks
this matching from being considered in the future.  It's also possible
for the CAS to provide clauses that block other similar matchings
(e.g., matchings generated via an automorphism of the graph) and we
showed that this was beneficial to the performance of the
solver~\cite{zulkoski2017combining}.

\paragraph{Combinatorial design theory.}
Exhaustive searches for Williamson matrices had
been performed in all odd orders
$n\leq59$~\cite{holzmann2008williamson} and all even orders
$n\leq18$~\cite{kotsireas2006constructions} prior to our work.  These searches
discovered that Williamson matrices don't exist in the orders $n=35$,
$47$, $53$, and $59$, but exist in all other orders that were searched.
Using \MC\ we were able to provide exhaustive searches for all orders
$n\leq70$ divisible by~$2$ or~$3$ (finding over 100,000 new sets of
Williamson matrices)~\cite{bright2019applying,bright2018satcas} and
verified the counterexample $n=35$~\cite{bright2016mathcheck}.

Using arithmetic circuits it is possible to generate a SAT instance
that specifies that Williamson matrices exist in order~$n$.  However,
this approach was only able to find Williamson matrices for orders up
to $n=30$.  To scale up to $n=70$ we found that it was essential to
use a divide-and-conquer approach and use CAS functionality in both
the divide and conquer phases.

First, we give an overview of the divide phase.  The first property that is useful in
this regard is the fact that Williamson matrices satisfy
\[ \s(A)^2+\s(B)^2+\s(C)^2+\s(D)^2 = 4n \]
where $\s(X)$ denotes the rowsum of the first row of~$X$. (We
associate a circulant matrix $X$ with the sequence formed by its first
row.)  We use a computer algebra system to solve the equation
$x^2+y^2+u^2+v^2=4n$ in integers and each solution provides one
subproblem, namely, the subproblem of finding a set of Williamson
matrices of order~$n$ with rowsums $(x,y,u,v)$.  This typically splits
each order into a few subproblems; to further divide the problem we
use \emph{sequence compression}~\cite{djokovic2015compression}.

For concreteness, suppose $n$ is even and $n=2m$.  Then Williamson
matrices can be ``compressed'' into matrices
of order $m$ by adding together the entries that are separated by
exactly $m$ entries in each row.  For example, the compression of the
row $[1,-1,1,1,-1,-1]$ is $[2,-2,0]$.  We generate further subproblems
using compressions; each subproblem corresponds to finding a set of
Williamson matrices that compresses to a given $\{\pm2,0\}$-sequence.
The reason this method of dividing is so effective is because there
are strong filtering theorems that a CAS can use to determine that
most subproblems of this form are unsatisfiable without even using a
SAT solver.  An example of these filtering theorems is described below
(in the context of the conquer phase).

Second, we give an overview of the conquer phase.  In this phase a SAT solver
receives a number of independent SAT instances that encode one subproblem
that was generated in the divide phase.  Using an off-the-shelf SAT solver
in this stage allowed us to scale to order $n=35$ (and in particular
verify that $35$ is the smallest counterexample of the Williamson
conjecture~\cite{bright2016mathcheck}).  However, using a CAS in the
conquer phase was found to be orders of magnitude faster
and allowed us to scale to $n=70$.

The reason that using a CAS produces such a dramatic improvement is
because it allows the usage of filtering theorems that are very strong%
---but the theorems cannot easily be directly encoded into Boolean
logic.  As an example, it is known that if $A=[a_0,\dotsc,a_{n-1}]$ is
the first row of a Williamson matrix then the bound
\[ \abs[\bigg]{\sum_{j=0}^{n-1}a_j\exp\paren[\big]{2\pi\sqrt{-1}jk/n}}^2 \leq 4n \]
holds for all integers $k$.  This is a very strict bound that
the vast majority of~$A$ will fail to satisfy for some~$k$.
Furthermore, it is very efficient to test because the
values on the left form the \emph{power spectral density} of~$A$
and can be quickly computed using a fast Fourier transform
(e.g., using the \textsc{DFT} and \textsc{PowerSpectrum} functions of the 
computer algebra system Maple).

In the conquer phase the SAT solver proceeds as normal until
a partial satisfying assignment of the propositional constraints
are such that the values of~$A$ can be determined.
At this point~$A$ is passed to
a CAS which computes its power spectrum.  If the power spectral
density bound is violated then a conflict clause is returned to the
SAT solver that blocks this~$A$ from being considered in the future.
The same condition is also checked with $B$, $C$, and $D$.

\paragraph{Number theory.}
All complex Golay pairs up to length $n=19$ were enumerated
in~\cite{craigen2002complex} and it was conjectured (based on
their partial search and patterns that they noticed)
that such pairs do not exist for $n=23$ and~$28$.
This was proven in~\cite{fiedler2013small} where a complete enumeration was
performed up to $n=28$. However, this result had never been
independently verified.  Using \MC\ we performed the first independent
verification of this result~\cite{bright2018complex} by explicitly
finding all complex Golay pairs for $n\leq28$, and further provided a
complete enumeration of all inequivalent complex Golay pairs up to~$28$.

At first it is not even obvious that a search for complex Golay pairs
of length~$n$ could be translated into SAT, since there are an infinite
number of~$z$ on the unit circle.  In fact, using arithmetic circuits
and other properties of complex Golay pairs it is possible to generate
a SAT instance that specifies that complex Golay pairs exist in
length~$n$.  However, in our experience these instances could only be
solved up to $n=16$ before incorporating a CAS\@.  Similar to the
previous case study we employ a divide and conquer approach and use
CAS functionality in both phases.

In the divide phase we perform a search for all possible
polynomials~$A$ that could appear as a member of a complex Golay pair.
A number of properties that~$A$ must satisfy are used as filtering
criteria, the main one being that $\abs{A(z)}^2\leq 2n$ for all $z$ on
the unit circle.  To test this bound we find the maximum of the
non-linear function $\abs{A(z)}^2$ for $z$ on the unit circle; for
example, this can be done with the Maple command \textsc{NLPSolve}.

In the conquer phase we solve a SAT instance for each possible~$A$
that was found in the dividing phase; a satisfying assignment of such
an instance will produce a~$B$ such that $(A,B)$ form a complex Golay
pair.  To do this we use the relationship $N_A+N_B=[2n,0,\dotsc,0]$
where $N_X$ is the \emph{autocorrelation} of the coefficients of $X$.
For example, $N_B$ can be computed with the Maple command
\textsc{AutoCorrelation} once the coefficients of~$B$ are known.  An
important optimization is that most values of $N_B$ can be computed
with only partial knowledge of $B$; this allows one to learn shorter
conflict clauses based on only a partial assignment of the SAT
instance.

\section{Best matrices}\label{sec:best}

We now apply our experience using \MC\ on the three case studies
described in Section~\ref{sec:sat+cas} to a new problem, namely,
the problem of finding best matrices from combinatorial design theory.
Best matrices are similar to Williamson matrices but exist in fewer
orders; in fact, if best matrices exist in order~$n$ then~$n$ must be
of the form $r^2+r+1$.  The best known
result~\cite{djokovic2018goethals} is that best matrices exist for all
$r\leq6$ and we use \MC\ to extend this result to $r\leq7$.
It is unknown if best matrices exist for any $r\geq8$.

\subsection{Background}\label{subsec:background}

Let $X$ be a square matrix of order $n$.  Recall that $X$ is
\emph{symmetric} if $x_{i,j}=x_{j,i}$ for all indices $(i,j)$,
\emph{skew} if $x_{i,j}=-x_{j,i}$ for all indices $i\neq j$
and $x_{i,i}=1$,
and \emph{circulant} if $x_{i,j}=x_{i+1,j+1}$ for all indices
reduced mod $n$.

Four matrices $A$, $B$, $C$, $D$ of order $n$
with $\pm1$ entries and positive diagonal entries
are \emph{best matrices} if they satisfy
the following three conditions:
\begin{enumerate}
\item[(1)] $A$, $B$, and $C$ are skew and $D$ is symmetric.
\item[(2)] $A$, $B$, $C$, and $D$ are pairwise commutative.
\item[(3)] $AA^T+BB^T+CC^T+DD^T$ is the scalar matrix $4nI$.
\end{enumerate}

An example of best matrices of order three
(where ``$+$'' denotes $1$ and ``$-$'' denotes $-1$) are
\[ A=B=C=\begin{bmatrix}
+ & - & + \\
+ & + & - \\
- & + & +
\end{bmatrix} \qquad D = \begin{bmatrix}
+ & + & + \\
+ & + & + \\
+ & + & +
\end{bmatrix} . \]
In this paper we will only consider \emph{circulant} best matrices in which
case condition~(2) is always satisfied.  Furthermore, condition~(1) is easy to enforce
since, for example, once the first half of the entries in the first row
are known they uniquely determine the values of the entries in the second half.
This still leaves an enormous search space, however.  Since there are
$(n-1)/2$ undetermined entries in each matrix a naive brute-force search
would check $2^{4(n-1)/2}=4^{n-1}$ quadruples---making the search space
for best matrices of order~$57$ about a quarter of a billion times
larger than the search space for best matrices of order~$43$
(the previous largest best matrices known).
Nevertheless, we were successful in our search for best matrices of order~$57$
by employing a number of powerful filtering theorems and using SAT solvers
to search the remaining space.

\subsection{Equivalence operations}\label{subsec:equivalence}

There are three operations on best matrices $A$, $B$, $C$, $D$ that can be used
to produce a new equivalent set of best matrices:
\begin{enumerate}
\item Reorder $A$, $B$, and $C$ in any way.
\item Apply the operation $f(i)\coloneqq-i\bmod n$ to the indices of
the first row of $A$, $B$, or~$C$.  (Since $D$ is symmetric such an operation
has no effect on it.)
\item Apply an automorphism of the cyclic group $\Z_n$ to the indices of the first rows
of $A$, $B$, $C$, and $D$ simultaneously.
\end{enumerate}

Such equivalence operations are well-known~\cite{djokovic2018goethals}.
The ``cyclic shift'' operation is sometimes also considered an equivalence operation
but we did not use it as it generally disturbs the symmetry and anti-symmetry
of the matrices.

\subsection{Divide phase}\label{subsec:divide}

Our aim in this phase is to split the problem of finding best matrices of given order~$n$
into subproblems such that each subproblem is easy enough to be solved with
a SAT solver (coupled with a CAS).

For concreteness we will focus on the case
$n=57$ and use the fact that $57=3\cdot19$ which allows us to ``3-compress''
the rows of best matrices to form compressed best matrices of order~$19$.
If $X$ is a sequence of length $57$ then its compression $\bar X$
is a sequence of length $19$ such that its $k$th entry is
\[ \bar x_k\coloneqq x_k + x_{k+19} + x_{k+2\cdot19} \qquad\text{for $0\leq k<19$}. \]
The reason why compression is so important is because of the following
\emph{PSD equality} that the compressions of best matrices must satisfy~\cite{djokovic2015compression}:
\[ \PSD_{\A}(k)+\PSD_{\B}(k)+\PSD_{\C}(k)+\PSD_{\D}(k) = 4n \qquad\text{for all $k$}. \tag{$*$}\label{eq:psd} \]
Here $\PSD_{\bar X}$ denotes the \emph{power spectral density} of $\bar X$ defined by
\[ \PSD_{\bar X}(k) \coloneqq \abs[\bigg]{\sum_{j=0}^{18}\bar x_j\exp\paren[\big]{2\pi\sqrt{-1}jk/19}}^2 . \] 
Our implementation (see Section~\ref{sec:implementation} for details) finds $15{,}178$ inequivalent possible
quadruples $(\A,\B,\C,\D)$ that satisfy the necessary relationship~\eqref{eq:psd} for $n=57$.
For each quadruple we generate a SAT instance with the $2n-2$ variables $\{\,a_i,b_i,c_i,d_i\,:\,1\leq i<(n+1)/2\,\}$.
The remaining entries are determined via the relationships
\[a_k = -a_{n-k},\quad b_k = -b_{n-k},\quad c_k = -c_{n-k},\quad d_k = d_{n-k} \qquad\text{for $k\neq0$.}\]
For clarity we will use variables with indices greater than $(n+1)/2$ with the understanding they refer to
variables in our SAT instance using these relationships.  Variables will be assigned \emph{true}
when they represent the entry~$1$ and \emph{false} when they represent the entry~$-1$
(by abuse of notation we use the same variable name for both
but it will be clear from context if
the variable is an integer or a Boolean).

Each of the $15{,}178$ possible compressions will specify a single independent SAT subproblem.
This is achieved by encoding the compression constraints in conjunctive normal form.
Because the sum of three $\pm1$ entries must be
$\pm3$ or~$\pm1$ these constraints come in four forms.

The first form is when $a_k+a_{k+19}+a_{k+2\cdot19}=3$.  In this case, we
add the cube $a_k\land a_{k+19}\land a_{k+2\cdot19}$ to the SAT subproblem.
The second form is when $a_k+a_{k+19}+a_{k+2\cdot19}=1$.  In this case,
we add
\[ (a_k\lor a_{k+19})\land (a_k\lor a_{k+2\cdot19}) \land (a_{k+19}\lor a_{k+2\cdot19}) \land (\lnot a_k\lor\lnot a_{k+19}\lor\lnot a_{k+2\cdot19}) \]
to the SAT subproblem.  The cases with $-1$ and $-3$ are handled in the same way with the polarity
of the literals in the clauses reversed.  We also add similar clauses for the entries of $B$, $C$, and~$D$.

\subsection{Conquer phase}\label{subsec:conquer}

Our aim in this phase is to solve the subproblems generated in the
dividing phase.  To do this, we employ a SAT solver with a
programmatic interface~\cite{ganesh2012lynx} that allows it to learn
conflict clauses by querying a CAS.  A programmatic SAT solver is
simply a variant of DPLL($T$), the key difference being
that in the programmatic SAT context the~$T$ solver can be specialized
to individual formulas (like an advice string in a non-uniform
computation model), whereas DPLL($T$) was envisioned with the $T$
solver being a decision procedure for an entire theory.
  
The property that the
CAS checks is the uncompressed form of the PSD equality~\eqref{eq:psd}, namely,
\[ \PSD_A(k) + \PSD_B(k) + \PSD_C(k) + \PSD_D(k) = 4n \qquad\text{for all $k$}. \]
Note that although this condition can only be verified to hold once
all entries of $A$, $B$, $C$, $D$ are known, in many cases it can be
verified to \emph{not} hold with only partial information.  In
particular, since $\PSD$ values are non-negative we must have the
\emph{PSD criterion}
\[ \sum_{X\in S}\PSD_X(k) \leq 4n \]
where $S$ is any subset of $\{A,B,C,D\}$.

In particular, if a partial assignment specifies
enough entries such that the PSD criterion
is violated then a conflict clause is learned that tells the
SAT solver to avoid that partial assignment in the future.
An important optimization is to choose $S$ in the PSD criterion
to be as small as possible.  For example, if both $S=\{A,B\}$
and $S=\{C\}$ violate the PSD criterion we prefer the latter
because in that case we learn a shorter conflict clause.
In the latter case the learned clause would say that at least one
variable $\{\,c_i\,:\,0\leq i<n\,\}$ has to be assigned differently
to its current assignment.

Additionally, the entries of best matrices
can be shown to satisfy certain constraints
similar to constraints that
Williamson matrices~\cite{williamson1944hadamard}, good matrices~\cite{bright2019good},
and the coefficients of complex Golay pairs~\cite{bright2018complex} satisfy.
In the appendix we show that the entries of best matrices satisfy the relationship
$a_kb_kc_kd_ka_{2k}b_{2k}c_{2k} = -1$ for $k\neq0$ with indices reduced mod~$n$.
Because of the anti-symmetry of $A$, $B$, and $C$, when $k=n/3$ the product constraint
reduces to $d_{k}=1$ and in this case can be encoded as a unit clause.
In general we encode the product constraint in SAT by breaking it up into the six constraints 
\[ x_0 = a_kb_k,\;\; x_1 = x_0c_k,\;\; x_2 = x_1d_k,\;\; x_3 = x_2a_{2k},\;\; x_4 = x_3b_{2k},\;\; x_5 = x_4c_{2k} \]
with $x_5=-1$, where the $x_i$ are new variables.  For example the first of these constraints
is represented in conjunctive normal form as
\[ (x_0\lor a_k\lor b_k)\land(\lnot x_0\lor \lnot a_k\lor b_k)\land(x_0\lor \lnot a_k\lor \lnot b_k)\land(\lnot x_0\lor a_k\lor \lnot b_k) \]
and the others are represented similarly.

\subsection{Example}

We now present an example of applying our method on a small order, namely, the order
$n=3$.
Let $(\A,\B,\C,\D)$ be the $3$-compression of the first rows of a set of best matrices
$(A,B,C,D)$ of order~$3$.

With the help of a CAS we find that the only possible solutions for
$(\A,\B,\C,\D)$ whose PSDs sum to 12 have $\bar{a}_0 = \pm1$, $\bar{b}_0 = \pm1$,
$\bar{c}_0 = \pm1$, and $\bar{d}_0 = \pm3$.
A CAS can also show that all $(\bar{a}_0, \bar{b}_0, \bar{c}_0, \bar{d}_0)$ are
equivalent to $(1, 1, 1, 3)$.

A SAT instance is generated with the variables $(a_i, b_i, c_i, d_i)$ for
$0\leq i<3$ and the following fifteen clauses:
\begin{gather*}
a_0 \lor a_1, \qquad a_0 \lor a_2, \qquad a_1 \lor a_2, \qquad \lnot a_0 \lor \lnot a_1 \lor \lnot a_2, \\
b_0 \lor b_1, \qquad b_0 \lor b_2, \qquad b_1 \lor b_2, \qquad \lnot b_0 \lor \lnot b_1 \lor \lnot b_2, \\
c_0 \lor c_1, \qquad c_0 \lor c_2, \qquad c_1 \lor c_2, \qquad \lnot c_0 \lor \lnot c_1 \lor \lnot c_2, \\
d_0, \qquad d_1, \qquad d_2 .
\end{gather*}
Finally, if the SAT solver finds a partial satisfying assignment of these
clauses such that $\sum_{X\in S}\PSD_X(k) > 4n$ for some subset~$S$ of $\{A,B,C,D\}$
and integer~$k$ then a conflict clause is programmatically learned that blocks
the variables that appear in~$S$ from being assigned the way they are in the partial
satisfying assignment.

In this case the only solutions are equivalent to those that assign the
variables $\{a_1, b_1, c_1\}$ to false and the other variables to true.
This solution leads to the example of best matrices given in Section~\ref{subsec:background}.

The product relationship $a_1b_1c_1d_1a_{2}b_{2}c_{2} = -1$
with $a_1=-a_2$, $b_1=-b_2$, and $c_1=-c_2$ reduces to $d_1=1$.
The unit clause $d_1$ has already been included in the SAT
instance, so this gives us no extra information in this case.
However, it can still be instructive to see how
this fact can be derived directly from the definition of best matrices.
(We focus on the case $n=3$; see the appendix for a general derivation.)

Consider the generating function $A(x)\coloneqq a_0+a_1x+a_2x^2$ where $x$
is an indeterminate variable with $x^3=1$ and similarly
for $B$, $C$, and $D$.  Note that $A(x)A(x^{-1})$ expands to
\[ (a_0^2+a_1^2+a_2^2) + (a_0a_1+a_1a_2+a_2a_0)x + (a_0a_2+a_1a_0+a_2a_1)x^2 . \]
This corresponds with
the entries of the first row of the circulant matrix $AA^T$, namely,
\[ \left[\begin{array}{c@{\qquad}c@{\qquad}c}
a_0^2+a_1^2+a_2^2 & a_0a_1+a_1a_2+a_2a_0 & a_0a_2+a_1a_0+a_2a_1 \\
\end{array}\right] . \]
Thus, it follows that for $A$, $B$, $C$, $D$ to be best matrices we must have
\[ A(x)A(x^{-1}) + B(x)B(x^{-1}) + C(x)C(x^{-1}) + D(x)D(x^{-1}) = 4\cdot 3 . \]
Using $a_1=-a_2$, $b_1=-b_2$, $c_1=-c_2$, $d_0=1$, and $d_1=d_2$,
we can derive
\begin{gather*}
A(x)A(x^{-1}) = 3-x-x^2, \\
B(x)B(x^{-1}) = 3-x-x^2, \\
C(x)C(x^{-1}) = 3-x-x^2, \\
D(x)D(x^{-1}) = 3+(1+2d_1)x+(1+2d_1)x^2.
\end{gather*}
Summing these together, we have
\[ 4\cdot3 + (-2+2d_1)x + (-2+2d_1) x^2 = 4\cdot 3 \]
from which it follows that $-2+2d_1=0$ and $d_1=1$.

\section{Implementation and results}\label{sec:implementation}

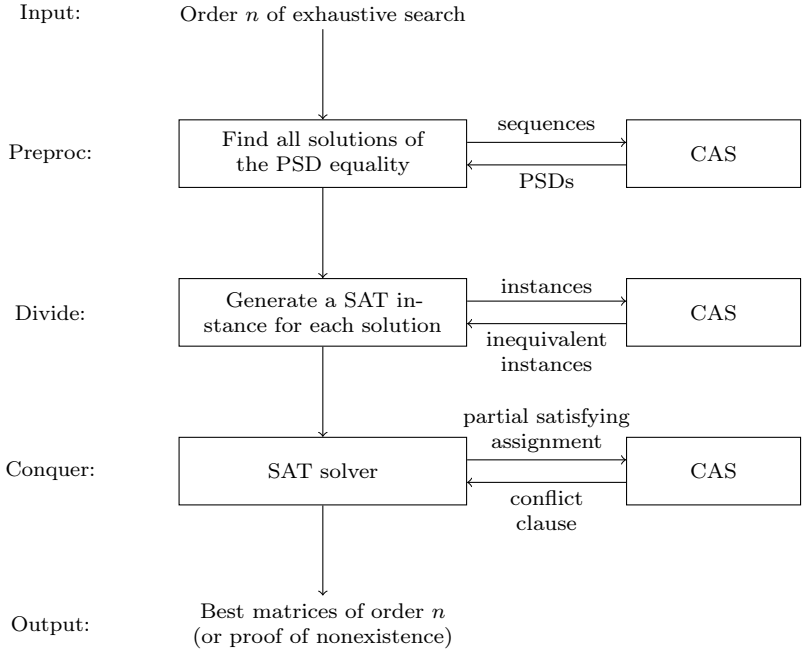
\begin{figure}
\begin{center}
\begin{tikzpicture}[align=center,node distance=4em]
\node(input){Order $n$ of exhaustive search};
\node[below=of input,text width=12em,minimum height=3em,rectangle,draw](divide){Find all solutions of the PSD equality};
\node[node distance=7em,right=of divide,text width=7em,minimum height=3em,rectangle,draw](cas){CAS};
\node[below=of divide,text width=12em,minimum height=3em,rectangle,draw](divide2){Generate a SAT instance for each solution};
\node[below=of cas,text width=7em,minimum height=3em,rectangle,draw](cas2){CAS};
\node[below=of divide2,text width=12em,minimum height=3em,rectangle,draw](sat){SAT solver};
\node[below=of cas2,text width=7em,minimum height=3em,rectangle,draw](cas3){CAS};
\node[below=of sat,text width=15em](output){Best matrices of order $n$ (or proof of nonexistence)};
\node[left of=input,node distance=12em](inputlabel){Input:};
\node[left of=divide,node distance=12em](preproclabel){Preproc:};
\node[left of=divide2,node distance=12em](preproclabel){Divide:};
\node[left of=sat,node distance=12em](conquerlabel){Conquer:};
\node[left of=output,node distance=12em](outputlabel){Output:};
\draw[->](input)--(divide);
\draw[->](divide)--(divide2);
\draw[->](divide2)--(sat);
\draw[->,transform canvas={yshift=0.5em}](divide)--node[above,text width=6em]{sequences}(cas);
\draw[<-,transform canvas={yshift=-0.5em}](divide)--node[below,text width=6em]{PSDs}(cas);
\draw[->,transform canvas={yshift=0.5em}](divide2)--node[above,text width=6em]{instances}(cas2);
\draw[<-,transform canvas={yshift=-0.5em}](divide2)--node[below,text width=6em]{inequivalent instances}(cas2);
\draw[->,transform canvas={yshift=0.5em}](sat)--node[above,text width=6em]{\clap{partial satisfying}\\assignment}(cas3);
\draw[<-,transform canvas={yshift=-0.5em}](sat)--node[below,text width=6em]{conflict clause}(cas3);
\draw[->](sat)--(output);
\end{tikzpicture}
\end{center}
\caption{A flowchart of our method for enumerating best matrices of order~$n$.
The PSD equality~\eqref{eq:psd} and divide step are described in Section~\ref{subsec:divide}
and the conquer step is described in Section~\ref{subsec:conquer}.
}\label{fig:flow}
\end{figure}

We implemented the divide and conquer phases described in Section~\ref{sec:best}
in our SAT+CAS system \MC.  Our code is available from our website
\href{https://uwaterloo.ca/mathcheck}{\nolinkurl{uwaterloo.ca/mathcheck}} along with more details for
the case studies described in Section~\ref{sec:sat+cas}.
A high-level flowchart of our method is shown in Figure~\ref{fig:flow}.

In the divide phase we wrote some custom C++ code to generate
all possible compressions of best matrices.  This code takes advantage
of the well-known fact (see~\cite{georgiou2001circulant}) that the rowsums of the first rows of $A$, $B$, $C$
must be $1$ and the squared rowsum of the first row of~$D$ must be
$4n-3$.  It follows that the rowsum of the first row of~$D$ is
$\pm(2r+1)$ where $n=r^2+r+1$.  In fact, the sign of $\s(D)$ is positive
when $r\equiv0,1\pmod{4}$ and negative otherwise (see appendix
for details).  Thus, for $r=7$ we have that $\s(D)=-15$.

We now employ a brute-force method to find all possibilities for the first
rows of best matrices of order~$n$.  
Taking into account the matrices are skew or symmetric there are $2^{(n-1)/2}$
possibilities for each of $A$, $B$, $C$, and $D$.
The majority of possibilities have a $\PSD$ value larger than $4n$ and
can therefore be ignored.  To further cut down on possibilities we also discard
possibilities that will lead to equivalent best matrices
using the equivalence operations of Section~\ref{subsec:equivalence}.
In particular, we apply operation~2 to the possibilities
for $B$ and~$C$ and operation~3 to the possibilities for $A$.

We then $3$-compress the possibilities for $n=57$,
finding $2748$ possibilities for~$\A$, $24{,}674$
possibilities for~$\B$ and~$\C$, and $7999$ possibilities
for~$\D$.  These possibilities now need to be joined into quadruples.
First, using brute-force we make a list of the possible pairs
$(\A,\B)$ and $(\C,\D)$; we find about $12$ million possibilities for the
former and $40$ million possibilities for the latter.
Then using the string
sorting and matching algorithm described in~\cite{kotsireas2009weighing}
we find all quadruples whose $\PSD$ values sum to $4n$.
After this step has completed we find $91{,}190$ possible quadruples
$(\A,\B,\C,\D)$ of which $15{,}178$ are inequivalent using the equivalence
operations of Section~\ref{subsec:equivalence}.
Each of these quadruples will form one independent subproblem (using
the SAT encoding described in Section~\ref{subsec:divide})
to be solved in the conquer phase.

For efficiency the $\PSD$ values were computed using the C library
FFTW~\cite{frigo2005design} that can very efficiently compute
discrete Fourier transforms.  Since FFTW uses floating-point arithmetic
we would only discard possibilities whose $\PSD$ values could
be shown to be larger than $4n+\epsilon$ where $\epsilon$ is larger
than the precision of the fast Fourier transform that was used.

In the conquer phase we solved our SAT instances using a
programmatic version of the SAT solver \textsc{MapleSAT}~\cite{liang2017empirical}.
The programmatic ``callback'' function was implemented as described in Section~\ref{subsec:conquer}
with a conflict clause being learnt whenever
enough of a partial assignment is known so that the PSD
criterion can be shown to be violated.
Again for efficiency we used the C library FFTW for computing
the $\PSD$ values.

As previously mentioned the orders of best matrices
must be of the form $r^2+r+1$ for $r\geq0$.
The case $r=7$ is currently the smallest open case and \MC\
is successfully able to solve this case.  For completeness, we
apply our method to smaller orders.
Orders of the form $r^2+r+1$ are prime for
$r\in\{1, 2, 3, 5, 6, 8\}$ and therefore do not
have a nontrivial compression factor.
Despite this, the cases $r=1$, $2$, $3$, $4$ can be be solved in under a second
using the method described in Section~\ref{sec:best}
with no compression (i.e., compression by~$1$).
Furthermore, the case $r=5$ can be solved in about $5$ seconds
and the case $r=6$ can be solved in about $50$ minutes.

We ran the case $r=7$ on a cluster of 64-bit Opteron 2.2GHz and Xeon 2.6GHz processors
running CentOS~6 using compression by a factor of~$3$.
In this case \MC\ requires about~$20$ minutes to perform the dividing
phase and about $162$ hours to perform the conquer phase.
These times measure the total amount of CPU time, though the conquer
phase took under an hour of real time when parallelized across~$200$ cores.

\begin{figure}
\centering\includegraphics[scale=0.33]{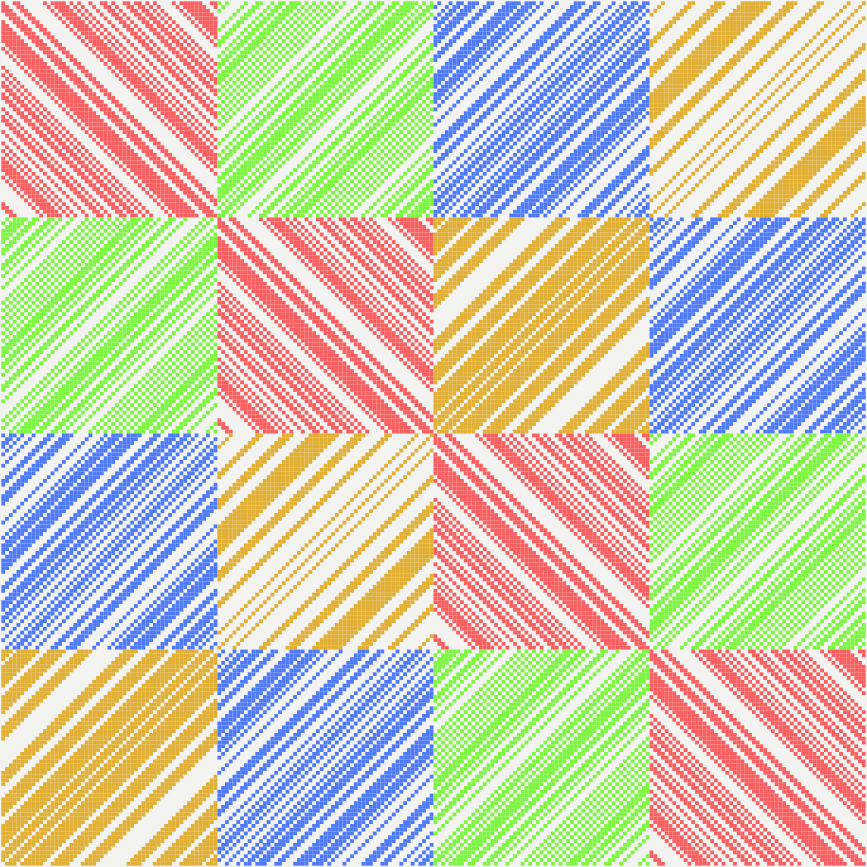}
\caption{A new skew Hadamard matrix of order $4\cdot57=228$
constructed using the Goethals--Seidel construction and
best matrices of order~$57$.  
The coloured entries represent $1$, the grey entries represent $-1$,
and each best matrix is coloured differently to more clearly show the
structure of the matrix.  (The matrices $B$, $C$, and~$D$ appear
reflected in the Goethals--Seidel construction.)}
\label{fig:hadamard}
\end{figure}

Three inequivalent sets of best matrices of order~$57$ were found
in the $r=7$ case.  We used the Goethals--Seidel construction~\cite{goethals1970skew}
to construct new skew Hadamard matrices of order $4\cdot57$ using these best matrices
and give one example in Figure~\ref{fig:hadamard}.
(Skew Hadamard matrices of order $4\cdot57$ have long been known~\cite{seberry1978skew}
but these are the first ones constructed using best matrices.)
Explicit representations of all the best matrices that we constructed
can be found on our website \href{https://uwaterloo.ca/mathcheck}{\nolinkurl{uwaterloo.ca/mathcheck}},
and explicit representations of the new best matrices are available in the appendix.

Let $B_r$ denote the number of inequivalent sets of best matrices of
order $r^2+r+1$.  Our results determine the value of $B_r$ for $r\leq7$:
\[ B_0 = 1, \;\; B_1 = 1, \;\; B_2 = 2, \;\; B_3 = 2, \;\; B_4 = 7, \;\; B_5 = 2, \;\; B_6 = 5, \;\; B_7 = 3 . \]
The value of $B_7$ is new, the value of $B_6$ was found in~\cite{djokovic2018goethals},
and the other values were given in~\cite{georgiou2001circulant}.
Our counts differ from those of~\cite{georgiou2001circulant} only because
that work did not use equivalence operation~2.
For example, for $r=4$ they find
twenty-one sets of best matrices but each of them is equivalent to one of the seven sets that we found.
The counts up to $r=5$ also appear
in~\cite{djokovic2009supplementary,koukouvinos2008skew}
but these works did not verify the counts.  To our knowledge we have performed
the first published verification.

\section{Conclusions and future work}\label{sec:conclusions}

We have described a ``SAT+CAS'' paradigm, building on DPLL($T$), that
is able to solve hard combinatorial problems that require \emph{both}
clever search routines (\`a la SAT) and efficient procedures for
complex mathematics outside the scope of traditional SMT theory
solvers (e.g., Fourier transforms in CAS).  As a demonstration of the
power and flexibility of the method we have outlined how it was used
to improve the state-of-the-art on three separate class of problems
from graph theory, number theory, and design theory, as well as its
application to construct new skew Hadamard matrices of order
$4\cdot57$.  The naive search space for such an object is
$2^{57\cdot56/2}\approx10^{480}$ which is totally impractical to
search using brute-force.  Instead, we use a number of mathematical
properties of best matrices to greatly constrain the search space.
However, it would still be too difficult to execute the search using either SAT
solvers or computer algebra systems in isolation: by themselves SAT solvers would
not be able to exploit the complex mathematical properties that are known and
computer algebra systems would not be able to exploit the efficient
search routines of SAT solvers.

We additionally find inspiration from the cube-and-conquer paradigm of
Heule et al.~\cite{heule2011cube}.  Since open problems (like finding
best matrices of order~$57$) typically have extremely large search
spaces they are usually not easy to solve using a sequential SAT
solver.  To deal with this we developed a method of dividing
instances into multiple independent subproblems.  In particular, we
divide the search for best matrices of order~$57$ into $15{,}178$
subproblems such that each subproblem can be solved in about a minute
using a SAT solver augmented with a domain-specific method of
generating conflicts.

Heule, Kullmann, and Marek~\cite{heule2017solving} point out that
there are essentially three kinds of solvers that are currently used
for solving large combinatorial problems: special-purpose solvers,
constraint satisfaction solvers, and SAT solvers.  We believe that
SAT+CAS solvers have now proven themselves as an effective way of
introducing the reasoning of special-purpose solvers and computer
algebra systems into SAT solving for hard problems from many areas of
mathematics.  Going forward, we expect that SAT+CAS solvers will
become essential for solving the largest combinatorial problems that
incorporate sophisticated mathematical properties.  For
example,~\cite{heule2017solving} points out that searching for finite
projective planes (a special kind of combinatorial design) has
currently only been done using special-purpose solvers.  These kinds
of problems are ripe for attack using the SAT+CAS paradigm.

\section*{Acknowledgments}

We thank the reviewers for their detailed comments that improved
the clarity of this article.

\bibliographystyle{spmpsci}
\bibliography{amai}

\input{appendix.tex}

\end{document}

%% file: appendix.tex
\section*{Appendix}

Let $A$, $B$, $C$, $D$ be a set of circulant best matrices of order $n=r^2+r+1$
(note that all numbers of this form are odd).
As described in Section~6 the rowsums of the first rows of $A$, $B$, $C$
are $1$ and the rowsum of the first row of~$D$ is $\pm(2r+1)$ where the
sign of $\s(D)$ is positive when $r\equiv0,1\pmod{4}$ and negative otherwise.

\begin{proof}
Since the matrix $A$ is skew we have $a_i+a_{n-i}=0$ for $i\neq0$.  Thus
\[ \s(A)=a_0+\sum_{i=1}^{(n-1)/2}(a_i+a_{n-i}) = 1 \]
and similarly for $B$ and $C$.  
Taking the relationship $AA^T + BB^T + CC^T + DD^T = 4nI$
and multiplying by the row vector of ones (on the left)
and the column vector of ones (on the right) we obtain
\[ \s(A)^2 + \s(B)^2 + \s(C)^2 + \s(D)^2 = 4n \]
and therefore $\s(D)^2=4n-3=(2r+1)^2$ and $\s(D)=s(2r+1)$ where $s=\pm1$.

Since $D$ is symmetric and $2d_i\equiv2\pmod{4}$
\[ \s(D) = 1 + 2\sum_{i=1}^{(n-1)/2}d_i \equiv n \pmod{4} . \]
Therefore $r^2+r+1\equiv s(2r+1)\pmod{4}$.  Since
\[ r^2+r+1 \equiv (-1)^{\floor{(r+1)/2}} \pmod{4} \qquad\text{and}\qquad 2r+1 \equiv (-1)^r \pmod{4} \]
we have $s=1$ when $r\equiv0,1\pmod{4}$ and $s=-1$ otherwise.
\qed
\end{proof}

As described in
Section~5.4 we have that the entries of these matrices satisfy the relationship
\[ a_k b_k c_k d_k a_{2k} b_{2k} c_{2k} = -1 \]
for $k\neq0$ with indices reduced mod $n$.

\begin{proof}

We can equivalently consider circulant best matrices to be polynomials
given by the generating function of the entries of their first rows.
In this formulation $A$, $B$, $C$, $D$ are polynomials
with $\pm1$ coefficients and of degree $n-1$
that satisfy
\[ A(x)A(x^{-1}) + B(x)B(x^{-1}) + C(x)C(x^{-1}) + D(x)D(x^{-1}) = 4n \tag{1} \]
modulo the ideal generated by $x^n-1$ (all computations will take place modulo this ideal).

Let $A_+$ denote the polynomial containing the terms of $A$ with positive coefficients
and let $\abs{A_+}$ denote the number of terms in $A_+$.
Then $A = 2A_+ - T$ where $T(x)\coloneqq\sum_{i=0}^{n-1}x^i$.
Since $x^iT=T$ we have $A_+T=\abs{A_+}T$ and $T^2=nT$.

Since $A$ is anti-symmetric (i.e., $A(x)+A(x^{-1})=2$) we have
$A(1)=1$ and $\abs{A_+}=(T(1)+A(1))/2=(n+1)/2$.  Furthermore,
\begin{align*}
A(x)A(x^{-1}) &= 2A-A^2 \\
&= 2(2A_+-T)-(2A_+-T)^2 \\
&= 4A_+ - 4A_+^2 - (2T - 4\abs{A_+}T + nT) \\
&= 4A_+ - 4A_+^2 + nT \tag{2}
\end{align*}
and similarly for $B$ and $C$.

Since $D$ is symmetric (i.e., $D(x)=D(x^{-1})$) we have
\begin{align*}
D(x)D(x^{-1}) &= (2D_+-T)^2 
= 4D_+^2 + (n-4\abs{D_+})T . \tag{3}
\end{align*}
By the symmetry of $D$ we have $D(x)=1+D'(x)+D'(x^{-1})$
where $D'(x)\coloneqq\sum_{i=1}^{(n-1)/2}d_ix^i$.
Then $\abs{D_+}=1+2\abs{D'_+}$
and thus $\abs{D_+}$ is odd.

Equating (1)--(3) and dividing by four we have
\[ A_+-A_+^2 + B_+-B_+^2 + C_+-C_+^2 + D_+^2 + (n-\abs{D_+})T = n . \tag{4} \]
Since $A_+=\sum_{a_i=1}x^i$ we have $A_+^2\equiv\sum_{a_i=1}x^{2i}\pmod{2}$ and~(4)
reduces to
\[ \sum_{a_i=1}(x^{2i}+x^i) + \sum_{b_i=1}(x^{2i}+x^i) + \sum_{c_i=1}(x^{2i}+x^i) + \sum_{d_i=1}x^{2i} \equiv 1 \pmod{2} \]
since both $n$ and $\abs{D_+}$ are odd.

Since $n$ is odd the congruence $i\equiv 2y\pmod{n}$ has exactly one solution
$0\leq y<n$ for each $0\leq i<n$.  Denoting this solution by $i/2$
we have
\[ \sum_{a_{i/2}=1}x^i + \sum_{a_i=1}x^i + \sum_{b_{i/2}=1}x^i + \sum_{b_i=1}x^i + \sum_{c_{i/2}=1}x^i + \sum_{c_i=1}x^i + \sum_{d_i=1}x^{2i} \equiv 1 \pmod{2} . \]
In other words, we have that the number of entries in
$\{a_{i/2},a_i,b_{i/2},b_i,c_{i/2},c_i,\allowbreak d_{i/2}\}$ that are positive
is $1$~(mod~$2$) for $i=0$
and $0$~(mod~$2$) for $i\neq0$.
Letting $k=i/2$ for $i\neq0$ this means $a_{k}a_{2k}b_kb_{2k}c_kc_{2k}d_k = -1$ as required.
\qed
\end{proof}

\begin{figure}
\begin{center}
\includegraphics[scale=0.39]{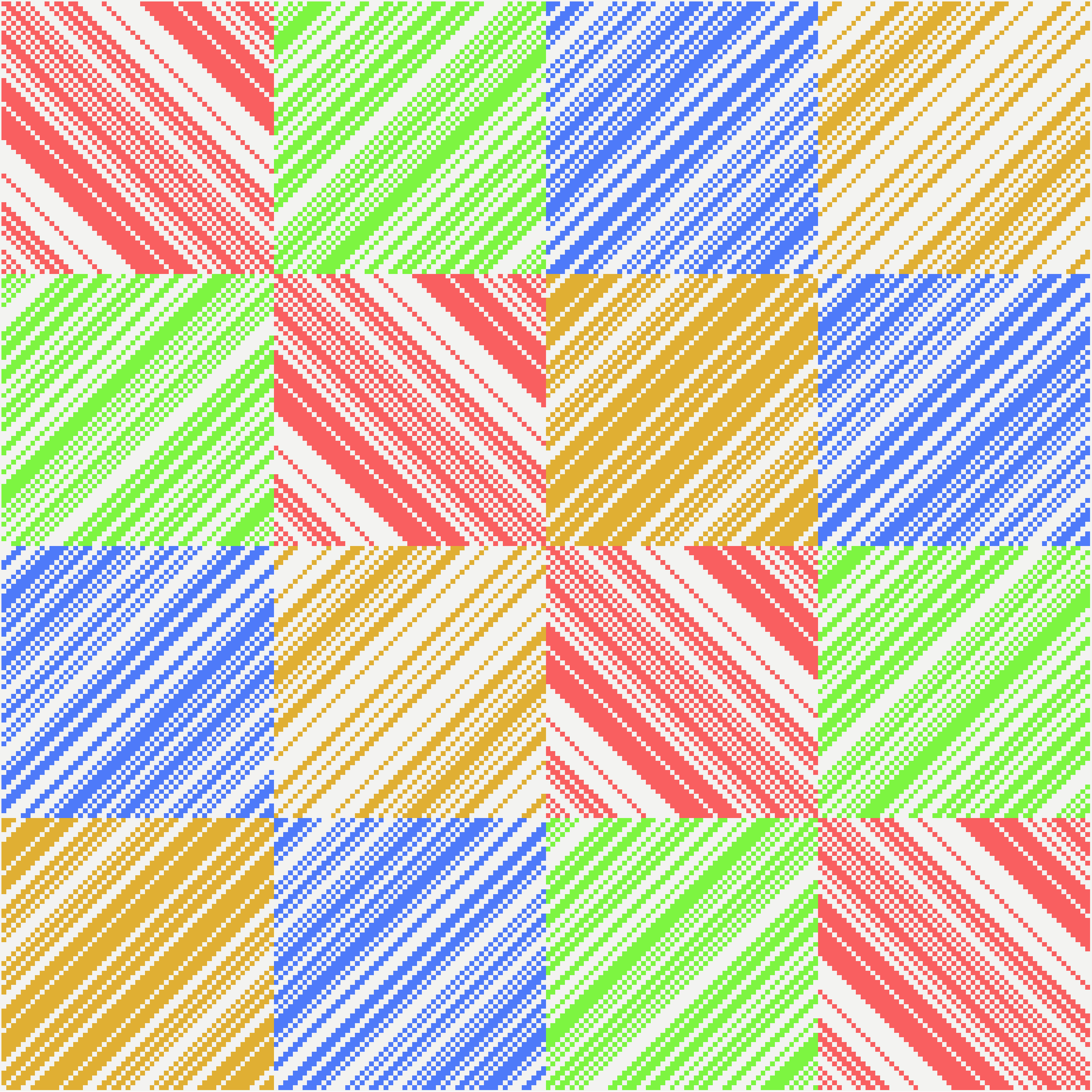}
\\[\baselineskip]
\includegraphics[scale=0.39]{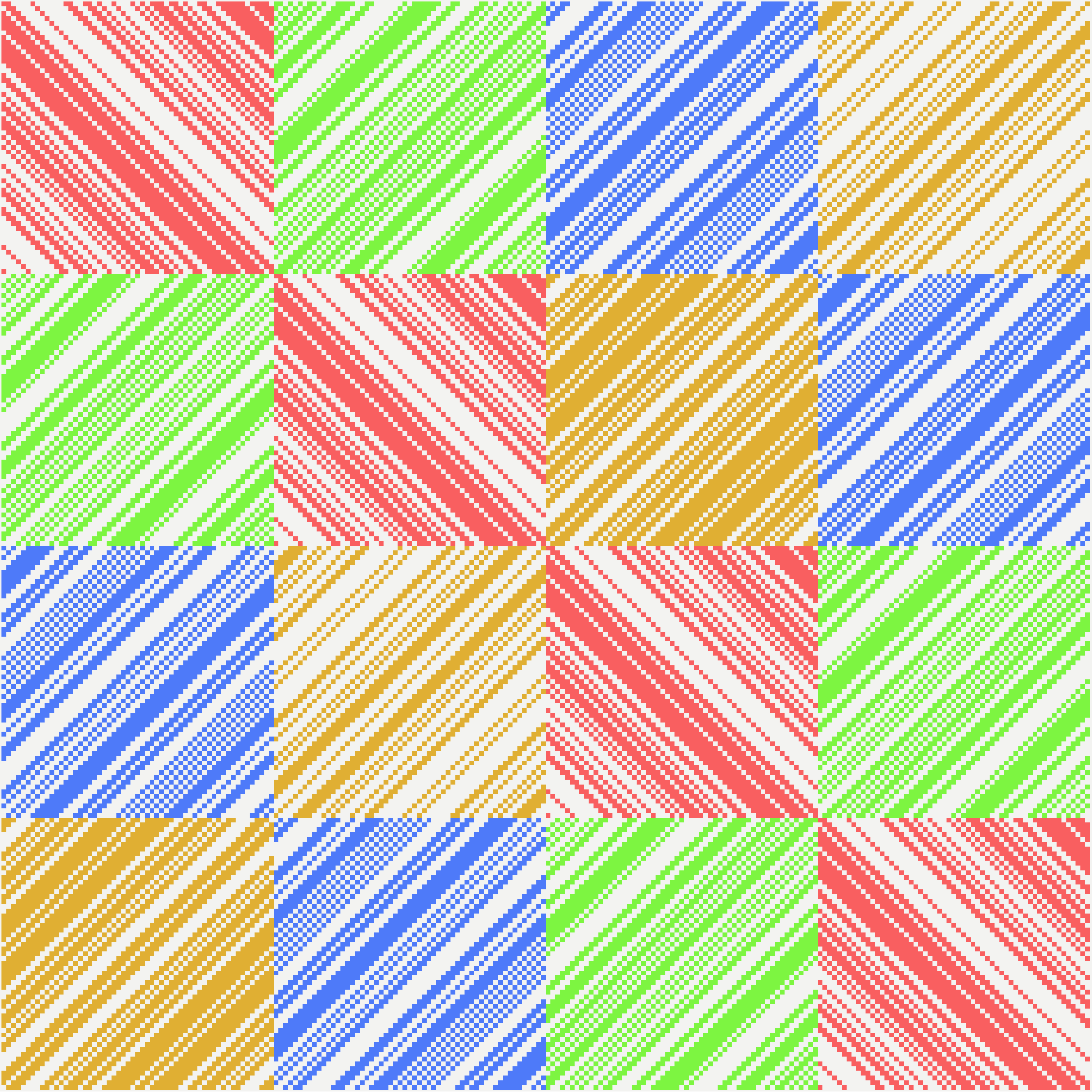}
\end{center}
\caption[LoF entry]{Two new skew Hadamard matrices of order $4\cdot57$ constructed using best matrices.
}
\label{fig:had3}
\end{figure}

One new skew Hadamard matrix that we constructed was given in Figure~\ref{fig:hadamard}
and the other two new skew Hadamard matrices are given in Figure~\ref{fig:had3}.
The first rows of the best matrices used to construct these skew Hadamard matrices are given here:

{\small
\microtypesetup{activate=false}\begin{center}
{\p\m\m\p\p\m\m\m\m\m\m\p\m\p\p\p\p\m\p\m\p\m\m\p\m\m\p\p\m\p\m\m\p\p\m\p\p\m\p\m\p\m\m\m\m\p\m\p\p\p\p\p\p\m\m\p\p} \\
{\p\m\p\m\p\p\p\p\m\p\p\m\m\p\p\m\p\m\m\p\m\m\m\p\m\p\m\p\m\p\m\p\m\p\m\p\p\p\m\p\p\m\p\m\m\p\p\m\m\p\m\m\m\m\p\m\p} \\
{\p\p\m\p\p\m\p\m\m\p\m\m\m\p\p\m\m\p\p\p\m\m\m\m\p\m\m\m\p\m\p\p\p\m\p\p\p\p\m\m\m\p\p\m\m\p\p\p\m\p\p\m\p\m\m\p\m} \\
{\p\m\m\m\p\m\m\m\m\m\m\p\p\m\m\m\m\m\p\p\p\m\m\m\m\p\p\p\p\p\p\p\p\m\m\m\m\p\p\p\m\m\m\m\m\p\p\m\m\m\m\m\m\p\m\m\m}
\\[\baselineskip]
{\p\m\m\p\p\m\m\m\m\m\m\p\m\p\p\p\p\m\p\m\p\m\m\p\m\m\p\p\m\p\m\m\p\p\m\p\p\m\p\m\p\m\m\m\m\p\m\p\p\p\p\p\p\m\m\p\p} \\
{\p\m\p\m\p\p\p\p\m\p\p\m\m\p\p\m\p\m\m\p\m\m\m\p\m\p\m\p\m\p\m\p\m\p\m\p\p\p\m\p\p\m\p\m\m\p\p\m\m\p\m\m\m\m\p\m\p} \\
{\p\p\m\p\p\m\p\m\m\p\m\m\m\p\p\m\m\p\p\p\m\m\m\m\p\m\m\m\p\m\p\p\p\m\p\p\p\p\m\m\m\p\p\m\m\p\p\p\m\p\p\m\p\m\m\p\m} \\
{\p\m\m\m\p\m\m\m\m\m\m\p\p\m\m\m\m\m\p\p\p\m\m\m\m\p\p\p\p\p\p\p\p\m\m\m\m\p\p\p\m\m\m\m\m\p\p\m\m\m\m\m\m\p\m\m\m}
\\[\baselineskip]
{\p\p\m\m\m\m\p\m\m\m\m\m\m\p\p\m\m\p\p\m\p\m\m\p\m\m\m\p\m\p\m\p\p\p\m\p\p\m\p\m\m\p\p\m\m\p\p\p\p\p\p\m\p\p\p\p\m} \\
{\p\m\m\p\m\p\m\p\m\m\p\m\p\p\m\m\m\m\p\p\m\m\p\p\p\p\p\p\m\p\m\m\m\m\m\m\p\p\m\m\p\p\p\p\m\m\p\m\p\p\m\p\m\p\m\p\p} \\
{\p\p\m\p\m\m\p\p\p\p\p\p\m\m\m\p\p\m\p\p\m\m\m\m\p\m\p\m\p\m\p\m\p\m\p\p\p\p\m\m\p\m\m\p\p\p\m\m\m\m\m\m\p\p\m\p\m} \\
{\p\m\m\m\p\p\p\p\m\m\p\m\p\m\m\m\p\m\m\p\p\m\m\m\m\m\m\p\m\m\p\m\m\m\m\m\m\p\p\m\m\p\m\m\m\p\m\p\m\m\p\p\p\p\m\m\m} \\
\end{center}}

